\documentclass[proceedings]{JHEP37}
\usepackage{eurosym}
\usepackage{amssymb}
\usepackage{amsfonts}
\usepackage{amsmath,epsfig}
\usepackage{graphicx}

\newbox\mybox

\newcommand\fverb{\setbox\mybox=\hbox\bgroup\verb}
\newcommand\fverbdo{\egroup\medskip\noindent\fbox{\unhbox\mybox}\ }
\newcommand\fverbit{\egroup\item[\fbox{\unhbox\mybox}]}
\conference{Regularized degenerate multi-solitons}
\abstract{We report complex $\mathcal{PT}$-symmetric multi-soliton solutions to the Korteweg de-Vries equation that asymptotically contain one-soliton solutions, with each of them
possessing the same amount of finite real energy. We demonstrate how these solutions originate from degenerate energy solutions of the Schr{\"{o}}dinger equation.
Technically this is achieved by the application of Darboux-Crum transformations involving Jordan states with suitable regularizing shifts. Alternatively they 
may be constructed from a limiting process within the context Hirota's direct method or on a nonlinear superposition 
obtained from multiple B{\"{a}}cklund transformations. The proposed procedure is completely generic and also applicable to other types of nonlinear integrable systems.  }

\title{Regularized degenerate multi-solitons}
\author{Francisco Correa$^{\dag,\ddag}$ and Andreas Fring$^\bullet$ \\
$\dag$ Instituto de Ciencias F{\'{\i}}sicas y Matem{\'{a}}ticas, Universidad
Austral de Chile, \\
$\,\,$ Casilla 567, Valdivia, Chile\\
$\ddag$ Institut f{\"{u}}r Theoretische Physik and Riemann Center for
Geometry and Physics, \\
$\,\,$ Leibniz Universit{\"{a}}t Hannover, Appelstra\ss e 2, 30167 Hannover,
Germany\\
$\bullet$ Department of Mathematics, City University London,\\
$\,\,$ Northampton Square, London EC1V 0HB, UK\\
E-mail: francisco.correa@itp.uni-hannover.de,a.fring@city.ac.uk}

\input{tcilatex}
\begin{document}

\section{Introduction}

Soliton solutions to nonlinear integrable wave equations play an important
role in nonlinear optics \cite{olver2012solitons}. The first successful
experiments to detect them have been carried out more than forty years ago 
\cite{bjorkholm}. A particularly important and structurally rich class of
solutions are multi-soliton solution which asymptotically behave as
individual one-soliton waves. This feature allows to view $N$-soliton
solutions as the scattering of $N$ single one-solitons with different
energies.

In analogy to von Neumann's avoided level crossing mechanism in quantum
mechanics \cite{vNW}, it is in general not possible to construct
multi-soliton solutions possessing asymptotically several one-solitons at
the same energy. The simple direct limit that equates two energies in the
expressions for the multi-solitons diverges in general. Some attempts have
been made in the past to overcome this problem. One may for instance
construct slightly modified multi-soliton solutions that allow for the
execution of a limiting process towards the same energy of some of the
multi-particle constituents \cite{kovalyov1,kovalyov2}. However, even though
the solutions found are mathematically permissible, they always possess
undesired singularities at certain points in space-time and have infinite
amounts of energy. These features make them non-physical objects.

Inspired by the success of $\mathcal{PT}$-symmetric quantum mechanics \cite%
{Bender:1998ke,Benderrev,Alirev}, many experiments have been carried out in
optical settings, exploiting the formal analogy between the Schr\"{o}dinger
and the Helmholtz equation. In particular, the existence of complex soliton
solutions in such a framework has recently been experimentally \cite%
{Muss,MatMakris,Guo} verified and it was shown \cite{CenFring} that such
type of solutions may posses real energies and lead to regular solutions
despite being complex. Here we will employ a similar idea and demonstrate
that they can be used to overcome the above mentioned infinite energy
problem related to degenerate multi-soliton solutions. Starting from a
quantum mechanical setting we show that the degeneracy is naturally
implemented by so-called Jordan states \cite{correa2015p} when Darboux-Crum
(DC) transforming \cite{darboux,crum,matasymp,matveevdarboux} degenerate
states of the Schr\"{o}dinger equation. Finiteness in the energy is achieved
by carefully selected complex $\mathcal{PT}$-symmetric shifts in the
dispersion terms.

Subsequently we show how such type of solutions are also obtainable from
other standard techniques of integrable systems. For Hirota's direct method 
\cite{Hirota} this can be achieved by reparameterizing known solutions such
that they will become suitable for a direct limiting process that leads to
degeneracy together with a fitting complexification that achieves the
regularization. For the other prominent scheme, the B\"{a}cklund
transformations we also demonstrate how the limit can be carried out on a
superposition of three solutions in a convergent manner.

Here we consider in detail one of the prototype nonlinear wave equations,
the Korteweg-de Vries (KdV) equation \cite{KdV}, for the complex field $%
u(x,t)$ 
\begin{equation}
u_{t}+6uu_{x}+u_{xxx}=0,\quad  \label{KdV}
\end{equation}%
depending on time $t$ and space $x$. This equation is known to arise from
standard functional variation from the Hamiltonian density 
\begin{equation}
\mathcal{H}(u,u_{x})=-u^{3}+\frac{1}{2}u_{x}^{2}.  \label{HH}
\end{equation}%
In general, for $\mathcal{PT}$-symmetric models the energy 
\begin{equation}
E=\int\nolimits_{-\infty }^{\infty }{\mathcal{H}}[u(x,t)]dx=\oint\nolimits_{%
\Gamma }{\mathcal{H}}[u(x,t)]\frac{du}{u_{x}},  \label{E}
\end{equation}%
remains real despite the fact that the Hamiltonian density is complex \cite%
{AFKdV}. The $\mathcal{PT}$-symmetry is realized as $\mathcal{PT}$: $%
x\rightarrow -x$, $t\rightarrow -t$, $i\rightarrow -i$, $u\rightarrow u$,
leaving (\ref{KdV}) invariant. As we will demonstrate below it is essential
to have complex contributions to $u$ in order to render the energy finite.

Our manuscript is organized as follows: In section 2 we discuss the general
mechanism that allows to implement degeneracies into Darboux-Crum
transformations. We show that degenerate states in the Schr\"{o}dinger
equation need to be replaced by Jordan states in order to obtain
nonvanishing and finite, up to singularities, solutions. Subsequently we
elaborate in detail on the novel features of degenerate two and three
soliton solutions and explain how the regularizing shifts need to be
implemented. In section 3 and 4 we explain how Hirota's direct method and
nonlinear superpositions obtained from four B\"{a}cklund transformations
need to be altered in order to allow for the construction of degenerate
complex multi-soliton solutions with finite energy. We state our conclusions
in section 5.

\section{Degenerate complex multi-soliton solutions from DC transformations}

\subsection{Darboux-Crum transformations, generalities}

The Darboux-Crum transformations \cite{darboux,crum,matasymp,matveevdarboux}
are well-known to generate covariantly an entire hierarchy of Schr\"{o}%
dinger equations to the same eigenvalue $E=-\lambda ^{2}~$in a recurrence
procedure. Following \cite{matveevdarboux} we recall here that in case of
degeneracy one has to replace the eigenstates of the Schr\"{o}dinger
equation by so-called \emph{Jordan states} $\Xi _{\lambda }^{(k)}$ defined
as solutions of the iterated Schr\"{o}dinger equation 
\begin{equation}
\hat{H}^{k+1}\Xi _{\lambda }^{(k)}=\left[ -\partial _{x}^{2}+V-E(\lambda )%
\right] ^{k+1}\Xi _{\lambda }^{(k)}=0,  \label{Jordan}
\end{equation}%
with potential $V$ and eigenvalue $E(\lambda )$ depending on the spectral
parameter $\lambda $. Thus for $k=0$ the corresponding Jordan state simply
becomes the eigenfunction of the Schr\"{o}dinger equation, that is $\Xi
_{\lambda }^{(0)}=\psi _{\lambda }~$or $\Xi _{\lambda }^{(0)}=\phi _{\lambda
}$ with $\phi _{\lambda }$ denoting the second fundamental solution to the
same eigenvalue $E(\lambda )$ obtainable via Liouville's formula $\phi
_{\lambda }(x)=\psi _{\lambda }(x)\dint\nolimits^{x}\left[ \psi _{\lambda
}(s)\right] ^{-2}ds$ from the first solution $\psi _{\lambda }$. The general
solution to (\ref{Jordan}) is easily seen to be%
\begin{equation}
\Xi _{\lambda }^{(k)}=\dsum\limits_{l=0}^{k}c_{l}\chi _{\lambda
}^{(l)}+\dsum\limits_{l=0}^{k}d_{l}\Omega _{\lambda }^{(l)},\qquad
c_{l},d_{l}\in \mathbb{R},  \label{sol}
\end{equation}%
with $\chi _{\lambda }^{(k)}:=\partial ^{k}\psi _{\lambda }/\partial E^{k}$
and $\Omega _{\lambda }^{(k)}:=\partial ^{k}\phi _{\lambda }/\partial E^{k}$%
. Some identities that will be useful below immediately arise from this.
Differentiating the Schr\"{o}dinger equation with respect to $E$ yields 
\begin{equation}
\hat{H}\left[ \chi _{\lambda }^{(1)}\right] =\psi _{\lambda },\qquad \text{%
and\qquad }\hat{H}\left[ \Omega _{\lambda }^{(1)}\right] =\phi _{\lambda },
\end{equation}%
which can be employed to derive 
\begin{equation}
W_{x}\left( \psi _{\lambda },\chi _{\lambda }^{(1)}\right) =-\psi _{\lambda
}^{2},\qquad \text{and\qquad }W_{x}\left( \phi _{\lambda },\Omega _{\lambda
}^{(1)}\right) =-\phi _{\lambda }^{2}.  \label{wx}
\end{equation}%
Here $W(\cdot )$ denotes the Wronskians $W(f,g)=fg_{x}-gf_{x}$.

Let us see how these states emerge naturally in degenerate
DC-transformations. With\ $E(\lambda )=-\lambda ^{2}$, the first iterative
step in this procedure is simply to note that the equation 
\begin{equation}
-\partial _{x}^{2}\psi _{\lambda }^{(1)}+V^{(1)}\psi _{\lambda
}^{(1)}=-\lambda ^{2}\psi _{\lambda }^{(1)},
\end{equation}%
with same eigenvalue as in (\ref{Jordan}) for $k=0$, but new potential%
\footnote{%
Here and in what follows we always understand $\left( \ln f\right) _{x}$ as
a short hand notation for $f_{x}/f$.} 
\begin{equation}
V^{(1)}=V-2\left( \ln \psi _{\lambda }^{(1)}\right) _{xx}
\end{equation}%
is solved by%
\begin{equation}
\psi _{\lambda }^{(1)}=\left\{ 
\begin{array}{ll}
D_{\psi _{\lambda _{1}}}(\psi _{\lambda })=W\left( \psi _{\lambda _{1}},\psi
_{\lambda }\right) \psi _{\lambda _{1}}^{-1}~~~~~ & \text{for~~}\lambda \neq
\lambda _{1} \\ 
~D_{\psi _{\lambda }}(\phi _{\lambda })=\psi _{\lambda }^{-1} & \text{for~~}%
\lambda =\lambda _{1}%
\end{array}%
\right. ,  \label{phi1}
\end{equation}%
where $D_{\psi }(\phi ):=\phi _{x}-\,(\psi _{x}/\psi )\phi $ is the Darboux
operator. The hierarchy of Schr\"{o}dinger equations is then obtained by
repeated application of these transformations. It is clear that a subsequent
iteration of the degenerate solution in (\ref{phi1}) will simply produce
again the potential $V$ and hence nothing novel. However, using the second
fundamental solution $\phi _{\lambda }^{(1)}$ $=\psi _{\lambda
}^{(1)}(x)\dint\nolimits^{x}\left[ \psi _{\lambda }^{(1)}(s)\right] ^{-2}ds$
to the level one equation yields something novel. In this case the new
potential becomes%
\begin{eqnarray}
V^{(2)} &=&V^{(1)}-2\left( \ln \phi _{\lambda }^{(1)}\right) _{xx}=V^{(1)}-2
\left[ \ln \left( \frac{1}{\psi _{\lambda }(x)}\dint\nolimits^{x}\left[ \psi
_{\lambda }(s)\right] ^{2}ds\right) \right] _{xx},  \label{V2} \\
&=&V-2\left[ \ln \left( \dint\nolimits^{x}\left[ \psi _{\lambda }(s)\right]
^{2}ds\right) \right] _{xx}=V-2\left[ \ln \left(
\dint\nolimits^{x}W_{s}\left( \psi _{\lambda },\chi _{\lambda }^{(1)}\right)
ds\right) \right] _{xx},~~~~  \notag \\
&=&V-2\left[ \ln \left[ W\left( \psi _{\lambda },\chi _{\lambda
}^{(1)}\right) \right] \right] _{xx},  \notag
\end{eqnarray}%
where we used identity (\ref{wx}) through which the Jordan states enter the
iteration procedure. The corresponding wave function to this potential is%
\begin{equation}
\psi _{\lambda }^{(2)}=D_{\phi _{\lambda }^{(1)}}\left[ D_{\psi _{\lambda
}}(\phi _{\lambda })\right] .
\end{equation}%
Proceeding in this way, the solutions to the hierarchy of equation 
\begin{equation}
-\partial _{x}^{2}\psi _{\lambda }^{(n)}+V^{(n)}\psi _{\lambda
}^{(n)}=-\lambda ^{2}\psi _{\lambda }^{(n)},\qquad n=0,1,2,\ldots 
\label{sch}
\end{equation}%
with potentials and wave functions%
\begin{eqnarray}
V^{(n)}(\lambda _{1},\ldots ,\lambda _{n}) &=&V^{(n-1)}-2\left( \ln \psi
_{\lambda }^{(n-1)}\right) _{xx}=V-2\ln \left[ W\left( \psi _{\lambda
_{1}},\ldots ,\psi _{\lambda _{n}}\right) \right] _{xx},  \label{Dar} \\
\psi _{\lambda }^{(n)}(\lambda _{1},\ldots ,\lambda _{n}) &=&D_{\psi
_{\lambda _{n}}^{(n-1)}}\left( \psi _{\lambda }^{(n-1)}\right)
=\dprod\limits_{k=1}^{n}D_{\psi _{\lambda _{k}}^{(k-1)}}\psi _{\lambda
}^{(0)}=\frac{W\left( \psi _{\lambda _{1}},\ldots ,\psi _{\lambda _{n}},\psi
_{\lambda }^{(0)}\right) }{W\left( \psi _{\lambda _{1}},\ldots ,\psi
_{\lambda _{n}}\right) },~~~  \notag
\end{eqnarray}%
respectively, for all $\lambda _{i}\neq \lambda _{j}$, $i,j=1,2,3,\ldots $
with $V^{(0)}=V$ and $\psi _{\lambda }^{(0)}=\psi _{\lambda }$ have to be
replaced by%
\begin{eqnarray}
V^{(n)}(\lambda ) &=&V^{(n-1)}-2\left( \ln \phi _{\lambda }^{(n-1)}\right)
_{xx}=V-2\ln \left[ W\left( \psi _{\lambda },\chi _{\lambda }^{(1)},\chi
_{\lambda }^{(2)},\ldots ,\chi _{\lambda }^{(n-1)}\right) \right]
_{xx},~~~~~~~ \\
\psi _{\lambda }^{(n)} &=&\dprod\limits_{k=1}^{n}D_{\phi _{\lambda
}^{(k-1)}}(D_{\psi _{\lambda }}(\phi _{\lambda }))=\frac{W\left( \psi
_{\lambda },\chi _{\lambda }^{(1)},\chi _{\lambda }^{(2)},\ldots ,\chi
_{\lambda }^{(n-1)}\right) }{W\left( \chi _{\lambda }^{(1)},\chi _{\lambda
}^{(2)},\ldots ,\chi _{\lambda }^{(n-1)}\right) }.  \notag
\end{eqnarray}%
Evidently we may also chose to have a partial degeneracy keeping some of the 
$\lambda _{i}$s different from each other, in which case we simply have to
replace consecutive $\psi _{\lambda _{i}}$ by Jordan states. For instance,
taking $\lambda _{1}\neq \lambda _{2}$ and $\lambda _{3}=\lambda
_{4}=\lambda _{5}=\lambda $ we obtain the potential 
\begin{equation}
V^{(5)}(\lambda _{1},\lambda _{2},\lambda ,\lambda ,\lambda )=V-2\ln \left[
W\left( \psi _{\lambda _{1}},\psi _{\lambda _{2}},\chi _{\lambda
}^{(1)},\chi _{\lambda }^{(2)},\chi _{\lambda }^{(3)}\right) \right] _{xx},
\end{equation}%
with either $\lambda =\lambda _{1}$ or $\lambda =\lambda _{2}$. Notice from (%
\ref{V2}) the sequence of Jordan states always has to accompanied by a $\chi
_{\lambda }^{(0)}=\psi _{\lambda }$. Let us now see how this procedure can
be employed in finding degenerate multi-soliton solutions by means of
inverse scattering.\ 

\subsection{Degenerate complex KdV multi-soliton solutions}

The different methods in integrable systems take various equivalent forms of
the KdV equation as their starting point. The Darboux-Crum transformation
exploits the fact that the central operator equation underlying all
integrable systems, the Lax equation $L_{t}=[M,L]$, may be written as a
compatibility equation between the two \emph{linear }equations 
\begin{equation}
L\psi =\lambda \psi \text{,\qquad and\qquad }\psi _{t}=M\psi \text{,~~~\ \ \
\ \ \ with }\psi =\psi (x,t,\lambda ),\lambda \in \mathbb{R}.  \label{lin}
\end{equation}%
For the KdV equation (\ref{KdV}) the operators are well-known to take on the
form%
\begin{equation}
L=-\partial _{x}^{2}-u,\text{\qquad and\qquad }M=-4\partial
_{x}^{3}-6u\partial _{x}-3u_{x}.
\end{equation}%
Thus $L$ becomes a Sturm-Liouville operator, such that the first equation in
(\ref{lin}) may be viewed as the Schr\"{o}dinger equation (\ref{sch}) with $%
L\equiv H$ being interpreted as a Hamiltonian operator. Considering now the
free theory with $u=0$ and taking the wave function in the form $\psi
(kx+\omega t)$, the second equation in (\ref{lin}) is solved by assuming the
nonlinear dispersion relation $4k^{2}+\omega =0$. For $\lambda =-\alpha
^{2}/4$ the two linear independent solutions to (\ref{lin}) are simply%
\begin{equation}
\psi _{\mu ,\alpha }(x,t)=\cosh \left[ \frac{1}{2}(\alpha x-\alpha ^{3}t+\mu
)\right] ,\text{\quad }\phi _{\mu ,a}(x,t)=\sinh \left[ \frac{1}{2}(\alpha
x-\alpha ^{3}t+\mu )\right] .  \label{fund}
\end{equation}%
We allowed here for a constant $\mu \in \mathbb{C}$ in the argument and
normalized the Wronskian as $W(\psi ,\phi )=\psi \phi _{x}-\psi _{x}\phi
=\alpha /2$. Suitably normalized, i.e. dropping overall factors, the first
Jordan states resulting from (\ref{fund}) are computed to%
\begin{eqnarray}
\chi _{\mu ,\alpha }^{(1)} &=&2\frac{\partial \psi _{\mu ,\alpha }}{\partial
\alpha }=\left( x-3\alpha ^{2}t\right) \phi _{\mu ,a},  \label{i1} \\
\chi _{\mu ,\alpha }^{(2)} &=&\alpha \left( x-3\alpha ^{2}t\right) ^{2}\psi
_{\mu ,\alpha }-2\left( x+3\alpha ^{2}t\right) \phi _{\mu ,a}, \\
\Omega _{\mu ,\alpha }^{(1)} &=&2\frac{\partial \phi _{\mu ,\alpha }}{%
\partial \alpha }=\left( x-3\alpha ^{2}t\right) \psi _{\mu ,\alpha }, \\
\Omega _{\mu ,\alpha }^{(2)} &=&a\left( x-3\alpha ^{2}t\right) ^{2}\phi
_{\mu ,a}-2\left( x+3\alpha ^{2}t\right) \psi _{\mu ,\alpha }.
\end{eqnarray}%
Using these explicit expressions the crucial identities (\ref{wx}) in the
above argument 
\begin{equation}
W_{x}\left( \psi _{\mu ,\alpha },\chi _{\mu ,\alpha }^{(1)}\right) =\alpha
\psi _{\mu ,\alpha }^{2},\qquad \text{and\qquad }W_{x}\left( \phi _{\mu
,\alpha },\Omega _{\mu ,\alpha }^{(1)}\right) =\alpha \phi _{\mu ,\alpha
}^{2},
\end{equation}%
are easily confirmed. We also verify%
\begin{equation}
\hat{H}\left[ \chi _{\mu ,\alpha }^{(1)}\right] =-\alpha \psi _{\mu ,\alpha
},~~\hat{H}\left[ \Omega _{\mu ,\alpha }^{(1)}\right] =-\alpha \phi _{\mu
,\alpha },~~\hat{H}^{2}\left[ \chi _{\mu ,\alpha }^{(2)}\right] =2\alpha
^{3}\psi _{\mu ,\alpha },~~\hat{H}^{2}\left[ \Omega _{\mu ,\alpha }^{(2)}%
\right] =2\alpha ^{3}\phi _{\mu ,\alpha },
\end{equation}%
which yield the defining relations for the Jordan states upon a subsequent
application of the energy shifted Hamiltonian $\hat{H}$ as defined in (\ref%
{Jordan}).

\subsubsection{Degenerate two-solitons}

To compute the degenerated two-soliton solution we use the above expressions
to evaluate the Wronskian $W(\psi _{\mu ,\alpha },\chi _{\mu ,\alpha }^{(1)})
$ involving one Jordan state. As indicated in (\ref{sol}) we may take the
constants $c_{l}$, $d_{l}$ different from zero, which we exploit here to
generate suitable regularizing shifts. First we compute%
\begin{eqnarray}
W\left[ \psi _{\mu ,\alpha },\chi _{\mu ,\alpha }^{(1)}\right]  &=&W\left[
\psi _{\mu ,\alpha },\left( x-3\alpha ^{2}t\right) \phi _{\mu ,\alpha }%
\right] =\left( x-3\alpha ^{2}t\right) W\left[ \psi _{\mu ,\alpha },\phi
_{\mu ,a}\right] +\psi _{\mu ,\alpha }\phi _{\mu ,\alpha }~~~~~~~  \notag \\
&=&\frac{1}{2}\left[ \alpha x-3\alpha ^{3}t+\sinh \left( \alpha x-\alpha
^{3}t+\mu \right) \right] ,
\end{eqnarray}%
where we used the identity (\ref{i1}) and the property of the Wronskian $%
W(f,gh)=W(f,g)h+fgh_{x}$. We note that one of the dispersion terms already
includes a shift $\mu $. Next we demand that also the dispersion term $%
\alpha x-3\alpha ^{3}t$ is shifted by a constant $\nu $, which is uniquely
obtained from 
\begin{equation}
W\left[ \psi _{\mu ,\alpha },\chi _{\mu ,\alpha }^{(1)}+\frac{\nu }{\alpha }%
\phi _{\mu ,\alpha }\right] =\frac{1}{2}\left[ \alpha x-3\alpha ^{3}t+\nu
+\sinh \left( \alpha x-\alpha ^{3}t+\mu \right) \right] .~~  \label{shift}
\end{equation}%
The degenerate two-soliton solution $u=2(\ln W)_{xx}$ resulting from (\ref%
{Dar}) and (\ref{shift}) reads 
\begin{equation}
u_{\mu ,\nu ;\alpha ,\alpha }(x,t)=\frac{2\alpha ^{2}\left[ \left( \alpha
x-3\alpha ^{3}t+\nu \right) \sinh \left( \alpha x-\alpha ^{3}t+\mu \right)
-2\cosh \left( \alpha x-\alpha ^{3}t+\mu \right) -2\right] }{\left[ \alpha
x-3\alpha ^{3}t+\nu +\sinh \left( \alpha x-\alpha ^{3}t+\mu \right) \right]
^{2}}.  \label{utwo}
\end{equation}%
This solution becomes singular when the Wronskian vanishes, which is always
the case for some specific $x$ and $t$ when $\nu ,\mu \in \mathbb{R}$.
However, for the $\mathcal{PT}$-symmetric choice $\nu =i\hat{\nu}$, $\mu =i%
\hat{\mu}$, $\hat{\nu},\hat{\mu}\in \mathbb{R}$ this solution becomes
regularized for a large range of choices for $\hat{\nu}$ and $\hat{\mu}$.
From 
\begin{equation}
W=\frac{1}{2}\left[ \cos \hat{\mu}\sinh \left( \alpha x-\alpha ^{2}t\right)
+\alpha x-3\alpha ^{3}t\right] +i\left[ \hat{\nu}+\sin \hat{\mu}\cosh
(\alpha x-\alpha ^{3}t)\right] ,
\end{equation}%
we observe that whenever $\hat{\nu}/\sin \hat{\mu}>-1$ the imaginary part of 
$W$ can not vanish and therefore $u_{\mu ,\nu ;\alpha ,\alpha }$ will be
regular in that regime of the shift parameters. Furthermore, we observe that 
$u_{\mu ,\nu ;\alpha ,\alpha }$ involves two different dispersion term $%
\alpha x-\alpha ^{3}t+\mu $ and $\alpha x-3\alpha ^{3}t+\nu $, each with a
separate shift. In the numerator the latter becomes negligible in the
asymptotic regimes where the degenerate two-soliton behaves as two single
solitons traveling at the same speed with one slightly decreasing and the
other with slightly increasing amplitude due to the time-dependent
pre-factor. In the intermediate regime, when the linear term $\alpha
x-3\alpha ^{3}t$ term in the numerator contributes, it produces a scattering
between the two one-solitons with the same energy. We depict this behaviour
in figure \ref{Deg2S}. In addition to the regularization, this entire
qualitative behaviour is due to the fact that our solutions are complex.

We observe that the larger and smaller amplitudes have exchanged their
relative position in the two asymptotic regimes with their mutual distance
kept constant. This is of course different from the standard nondegenerate
case where the solitons continuously approach each other before the
scattering event and separate afterwards. Again this is achieved through the
complexification of our solution. Here the scattering is governed by some
internal breatherlike structure as in confined to a certain region.

\FIGURE{ \epsfig{file=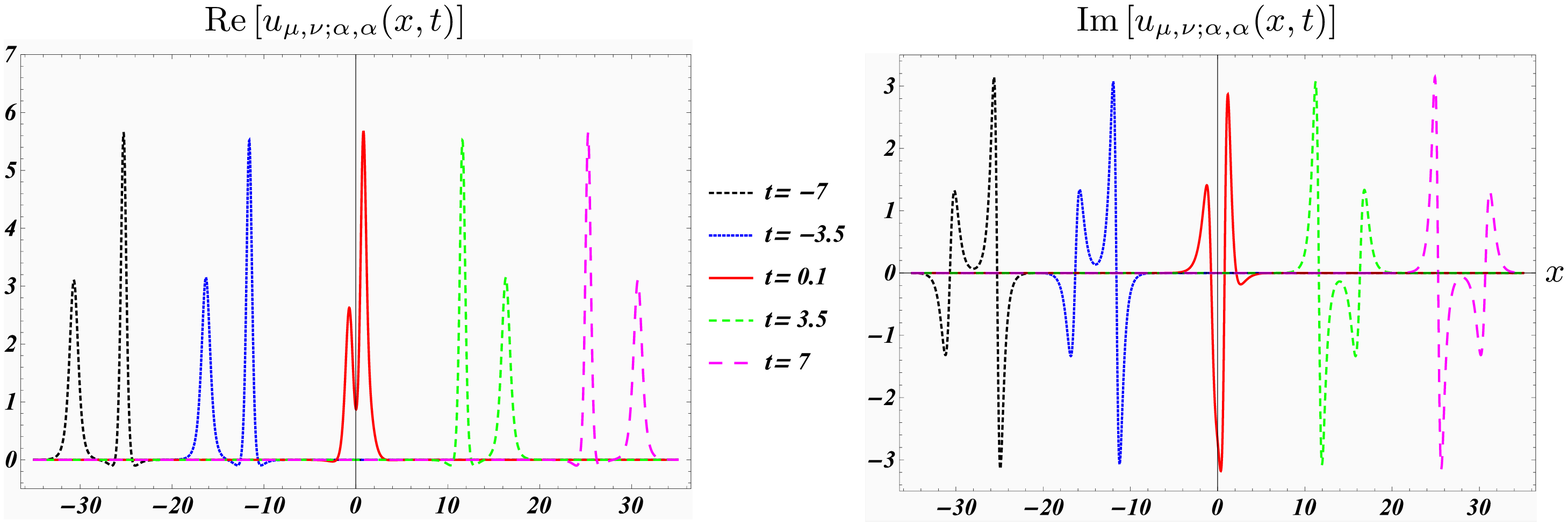,height=5.0cm} 
\caption{Degenerated KdV two-soliton compound solution with $\alpha=\beta=2$, $\mu=i \pi 3/5$ and $\nu=i \pi/5$ at different times.}
        \label{Deg2S}}

As demonstrated in figure \ref{Deg2Snu} this internal structure can be
manipulated by varying the shift $\nu $.

\FIGURE{ \epsfig{file=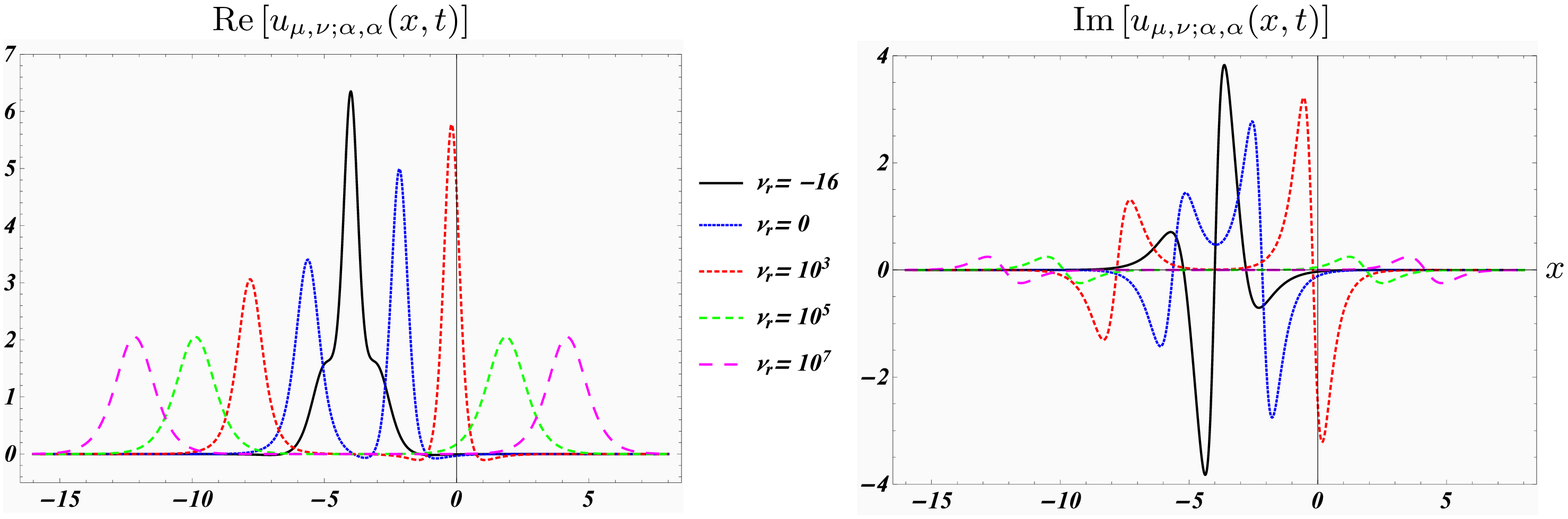,height=5.0cm} 
\caption{Degenerated KdV two-soliton compound solution with $\alpha=\beta=2$, $\nu=i \pi/5$ at fixed moment in time $t=-1$ and varying shift parameter $\nu=\nu_r + i \pi 3/5$.}
        \label{Deg2Snu}}

For a fixed instance in time we can employ $\nu $ to increase or decrease
the distance between the single soliton amplitudes and even find a value
such that the distance becomes zero. However, this value is in the
intermediate regime and as time evolves the two solitons will separate again
to some finite distance in the asymptotic regime.

Our interpretation is supported by the computation of the energies resulting
from (\ref{E}) with Hamiltonian density (\ref{HH}) for the solution $u_{\mu
,\nu ;\alpha ,\alpha }$. Numerically we find the finite real energies%
\begin{equation}
E_{\mu ,\nu ;\alpha ,\alpha }=\int\nolimits_{-\infty }^{\infty }{\mathcal{H}}%
[u_{\mu ,\nu ;\alpha ,\alpha },(u_{\mu ,\nu ;\alpha ,\alpha })_{x}]dx=-2%
\frac{\alpha ^{5}}{5}=2E_{\mu ;\alpha },
\end{equation}%
i.e. precisely twice the energy of the one-soliton $u_{\mu ;\alpha }$,
reported for instance in \cite{CenFring}.

In order to compare with various other methods it is useful to note that the
degenerate Wronskians may be obtained in several alternative ways. We
conclude this subsection by reporting how the expression for the Wronskian (%
\ref{shift}) can be derived by mean of a limiting process directly from the
two-soliton solution. This is seen from by starting from the defining
relation for the Jordan state $\chi ^{(1)}$ 
\begin{eqnarray}
W\left[ \psi _{\mu ,\alpha },\chi _{\mu ,\alpha }^{(1)}\right]
&=&2\lim_{\beta \rightarrow \alpha }\frac{\partial }{\partial \beta }W\left[
\psi _{\mu ,\alpha },\psi _{\mu ,\beta }\right]  \label{last} \\
&=&2\lim_{\beta \rightarrow \alpha }\lim_{h\rightarrow 0}W\left[ \psi _{\mu
,\alpha },\frac{\psi _{\mu ,\beta +h}-\psi _{\mu ,\beta }}{h}\right] \\
&=&2\lim_{h\rightarrow 0}W\left[ \psi _{\mu ,\alpha },\frac{\psi _{\mu
,\alpha +h}-\psi _{\mu ,\alpha }}{h}\right] \\
&=&2\lim_{h\rightarrow 0}\frac{1}{h}W\left[ \psi _{\mu ,\alpha },\psi _{\mu
,\alpha +h}\right] \\
&=&2\lim_{\beta \rightarrow \alpha }\frac{1}{\beta -\alpha }W\left[ \psi
_{\mu ,\alpha },\psi _{\mu ,\beta }\right] ,  \label{55}
\end{eqnarray}%
where in the last step we chose $h=\beta -\alpha $. The shift can now be
implemented by determining $\lambda $ from the limit of the expression%
\begin{eqnarray}
W\left[ \psi _{\mu +\lambda \nu ,\alpha },\psi _{\mu -\lambda \nu ,\beta }%
\right] &=&W\left[ \psi _{\mu ,\alpha },\psi _{\mu ,\beta }\right] \cosh
^{2}\left( \frac{\lambda \nu }{2}\right) -W\left[ \phi _{\mu ,\alpha },\phi
_{\mu ,\beta }\right] \sinh ^{2}\left( \frac{\lambda \nu }{2}\right) ~~~~~~~~
\\
&&+\frac{1}{2}\sinh \left( \lambda \nu \right) \left[ W\left[ \phi _{\mu
,\alpha },\psi _{\mu ,\beta }\right] -W\left[ \psi _{\mu ,\alpha },\phi
_{\mu ,\beta }\right] \right] .  \notag
\end{eqnarray}%
It it is obvious that for the limit (\ref{55}) of the shifted expression to
be finite we require $\lambda \sim (\alpha -\beta )$ with constant of
proportionality chosen in such a way that it yields $1/2\alpha $ in the
limit. Hence we obtain 
\begin{eqnarray}
W\left[ \psi _{\mu ,\alpha },\chi _{\mu ,\alpha }^{(1)}+\frac{\nu }{\alpha }%
\phi _{\mu ,\alpha }\right] &=&2\lim_{\beta \rightarrow \alpha }\frac{1}{%
\beta -\alpha }W\left[ \psi _{\mu +\frac{\alpha -\beta }{\alpha +\beta }\nu
,\alpha },\psi _{\mu -\frac{\alpha -\beta }{\alpha +\beta }\nu ,\beta }%
\right] ,  \label{shift2} \\
&=&2\lim_{\beta \rightarrow \alpha }W_{\beta }\left[ \psi _{\mu +\frac{%
\alpha -\beta }{\alpha +\beta }\nu ,\alpha },\psi _{\mu -\frac{\alpha -\beta 
}{\alpha +\beta }\nu ,\beta }\right] .~
\end{eqnarray}%
These identities will be useful below when we relate this approach to
Hirota's direct method.

\subsubsection{Degenerate three-solitons}

To find the degenerate three-soliton solution we may once again compute the
Wronskian, albeit now involving two Jordan states. As discussed in the
previous section, the expression for $W(\psi _{\mu ,\alpha },\chi _{\mu
,\alpha }^{(1)},\chi _{\mu ,\alpha }^{(2)})$ will inevitably lead to
solutions with infinite energy. Thus we will again exploit (\ref{sol}) with
nonvanishing constants $c_{l}$, $d_{l}$ to generate the regularizing $%
\mathcal{PT}$-symmetric shifts. A suitable unique choice is%
\begin{eqnarray}
&&W\left[ \psi _{\mu ,\alpha },\chi _{\mu ,\alpha }^{(1)}+\frac{\rho }{%
\alpha }\phi _{\mu ,\alpha },\chi _{\mu ,\alpha }^{(2)}+2\rho ~\Omega _{\mu
,\alpha }^{(1)}+\frac{2\nu -4\rho }{\alpha }\phi _{\mu ,\alpha }\right]
~~~~~~~\ \ ~~ \\
&=&\alpha \left[ 1+\left( \eta _{\rho ;\alpha }^{(3)}\right) ^{2}+\cosh
\left( \eta _{\mu ;\alpha }^{(1)}\right) \right] \sinh \left( \frac{\eta
_{\mu ;\alpha }^{(1)}}{2}\right) -\alpha \eta _{\nu ;\alpha }^{(9)}\cosh
\left( \frac{\eta _{\mu ;\alpha }^{(1)}}{2}\right) ,  \label{degthree}
\end{eqnarray}%
where we abbreviated the different dispersion terms as%
\begin{equation}
\eta _{\mu ;\alpha }^{(\lambda )}:=\alpha x-\lambda \alpha ^{3}t+\mu ,
\end{equation}%
Notice that we have now three different shifted dispersion terms $\eta _{\mu
;\alpha }^{(1)}$, $\eta _{\rho ;\alpha }^{(3)}$ and $\eta _{\nu ;\alpha
}^{(9)}$, where the first governs the asymptotic behaviour and the remaining
ones the additional structure in the intermediate regime. The solution $%
u_{\mu ,\nu ,\rho ;\alpha ,\alpha ,\alpha }=2(\ln W)_{xx}$ is depicted in
figure \ref{Deg3S}.

\FIGURE{ \epsfig{file=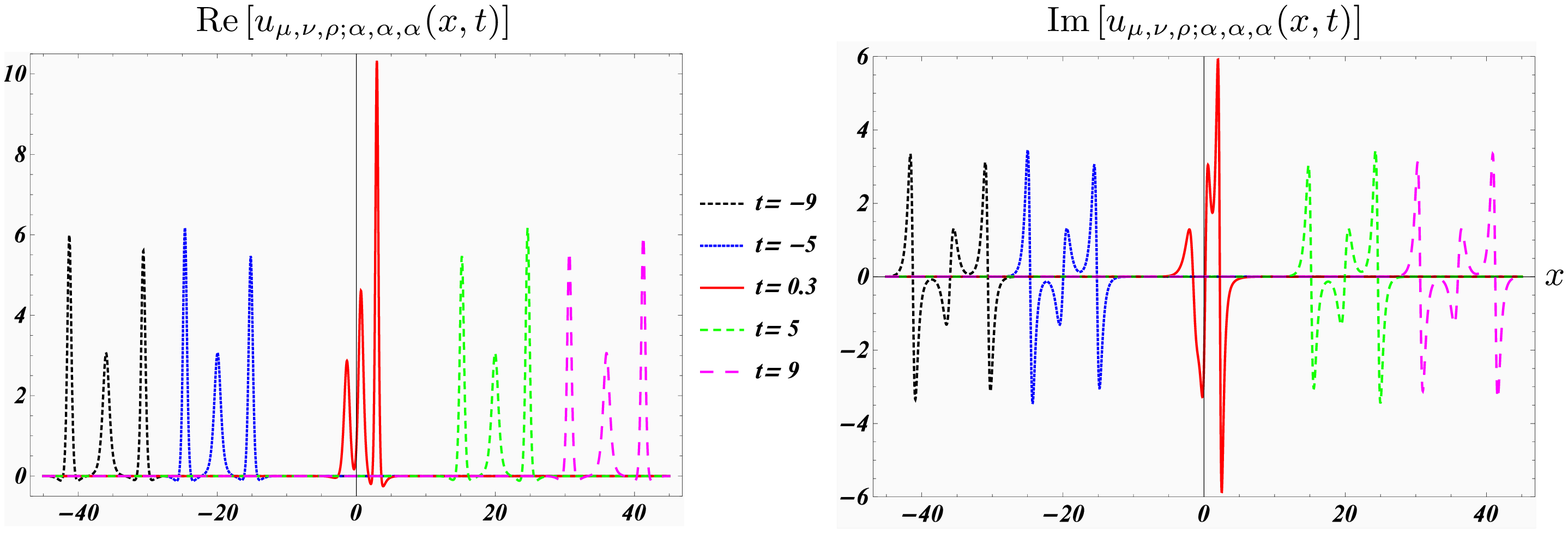,height=5.0cm} 
\caption{Degenerated KdV  three-soliton compound solution with $\alpha=\beta=\gamma=2$, $\mu=i \pi 3/5$,  $\nu=i \pi 3/10$ and $\rho=i \pi/10$.}
        \label{Deg3S}}

We observe that asymptotically we have three single one-solitons moving at
the same speed. They exchange their positions in the intermediate region
near the origin, when the linear terms in (\ref{degthree}) contribute.

For the general three-soliton solution we have also additional options
available, namely to produce the degeneracy only in two of the one-solitons
while keeping the remaining one at a different velocity. A suitable choice
that produces the desired shifts is%
\begin{eqnarray}
W\left[ \psi _{\mu ,\alpha },\chi _{\mu ,\alpha }^{(1)}+\frac{\nu }{\alpha }%
\phi _{\mu ,\alpha },\psi _{\rho ,\gamma }\right] &=&\left[ \frac{\alpha
^{2}+\gamma ^{2}}{8}\sinh \left( \eta _{\mu ;\alpha }^{(1)}\right) -\frac{%
\alpha ^{2}-\gamma ^{2}}{8}\eta _{\nu ;\alpha }^{(3)}\right] \cosh \left( 
\frac{\eta _{\rho ;\gamma }^{(1)}}{2}\right)  \notag \\
&&-\frac{\alpha \gamma }{2}\cosh ^{2}\left( \frac{\eta _{\mu ;\alpha }^{(1)}%
}{2}\right) \sinh \left( \frac{\eta _{\rho ;\gamma }^{(1)}}{2}\right) .
\end{eqnarray}%
We depict the corresponding KdV solution $u_{\mu ,\nu ,\rho ;\alpha ,\alpha
,\gamma }=2(\ln W)_{xx}$ in figure \ref{Deg21S}.

\FIGURE{ \epsfig{file=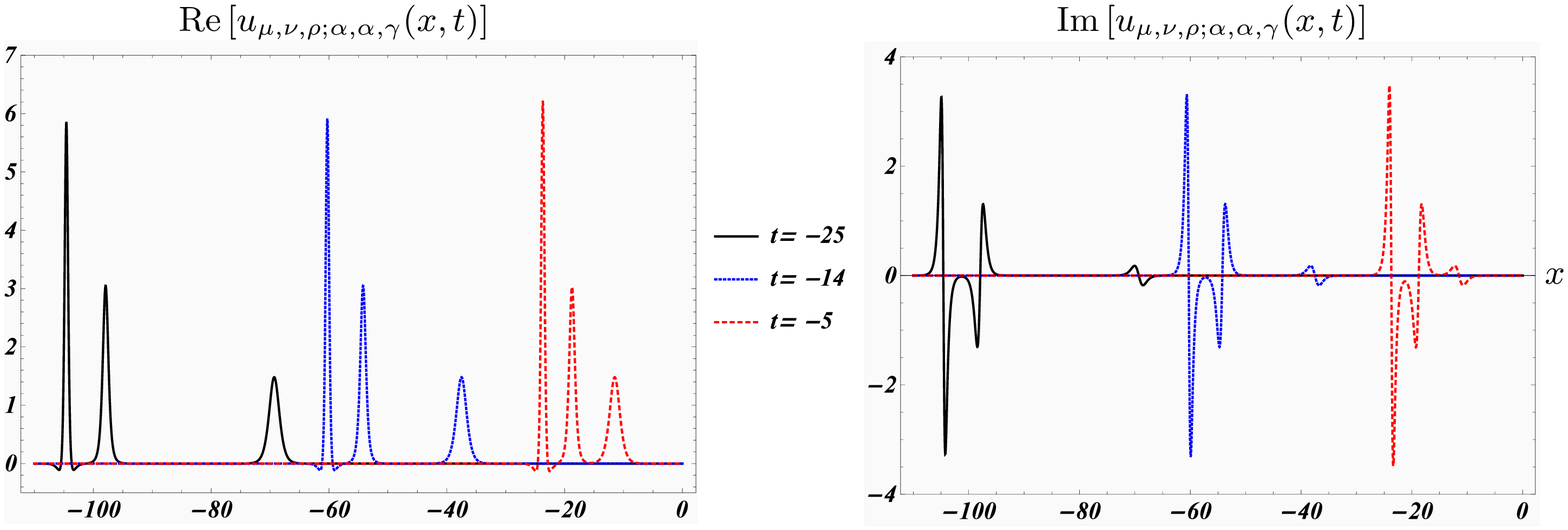,height=5.0cm} 
\caption{Degenerated KdV two-soliton compound solution scattering with a one-soliton with $\alpha=\beta=2$, $\gamma=1.7$, $\mu=i \pi 3/5$,  $\nu=i \pi 3/10$ and $\rho=i \pi/10$.}
       \label{Deg21S}}

We clearly observe that asymptotically we have a degenerated two-soliton and
a one-soliton solution with the faster two-soliton overtaking the slower
one-soliton.

Let us finish this section by reporting an alternative form of the
degenerate three-soliton solution suitable for a comparison with other
methods. We find

\begin{eqnarray}
&&W\left[ \psi _{\mu ,\alpha },\chi _{\mu ,\alpha }^{(1)}+\frac{\rho }{%
\alpha }\phi _{\mu ,\alpha },\chi _{\mu ,\alpha }^{(2)}+2\rho \Omega _{\mu
,\alpha }^{(1)}+\frac{2\nu -4\rho }{\alpha }\phi _{\mu ,\alpha }\right] \\
&=&16\alpha \lim_{\gamma ,\beta \rightarrow \alpha }\frac{1}{(\alpha -\beta
)(\alpha -\gamma )(\gamma -\beta )}W\left[ \psi _{\mu +f(\alpha ,\beta
,\gamma ),\alpha },\psi _{\mu +f(\beta ,\gamma ,\alpha ),\beta },\psi _{\mu
+f(\gamma ,\alpha ,\beta ),\gamma }\right] ~~~~~~~\ \ ~~ \\
&=&8\alpha \lim_{\gamma ,\beta \rightarrow \alpha }W_{\beta \gamma \gamma } 
\left[ \psi _{\mu +f(\alpha ,\beta ,\gamma ),\alpha },\psi _{\mu +f(\beta
,\gamma ,\alpha ),\beta },\psi _{\mu +f(\gamma ,\alpha ,\beta ),\gamma }%
\right] ,
\end{eqnarray}%
where we introduced the shift function%
\begin{equation}
f(x,y,z):=\frac{4}{9}\left[ \frac{x^{2}+yz}{(x+y)(x+z)}-2\frac{x(y^{2}+z^{2})%
}{(x+y)(x+z)(y+z)}\right] \nu +\frac{4}{3}\frac{x^{2}-yz}{(x+y)(x+z)}\rho .
\label{f}
\end{equation}%
It will be important below to note that the sum of all shifts adds up to
zero, $f(\alpha ,\beta ,\gamma )+f(\beta ,\gamma ,\alpha )+f(\gamma ,\alpha
,\beta )=0$.

Once again our interpretation is supported by the computation of the
corresponding energies. Numerically we find 
\begin{eqnarray}
E_{\mu ,\nu ,\rho ;\alpha ,\alpha ,\alpha } &=&\int\nolimits_{-\infty
}^{\infty }{\mathcal{H}}[u_{\mu ,\nu ,\rho ;\alpha ,\alpha ,\alpha },(u_{\mu
,\nu ,\rho ;\alpha ,\alpha ,\alpha })_{x}]dx=-3\frac{\alpha ^{5}}{5}=3E_{\mu
;\alpha }, \\
E_{\mu ,\nu ,\rho ;\alpha ,\alpha ,\gamma } &=&\int\nolimits_{-\infty
}^{\infty }{\mathcal{H}}[u_{\mu ,\nu ,\rho ;\alpha ,\alpha ,\gamma },(u_{\mu
,\nu ,\rho ;\alpha ,\alpha ,\gamma })_{x}]dx=-2\frac{\alpha ^{5}}{5}-\frac{%
\gamma ^{5}}{5}=2E_{\mu ;\alpha }+E_{\mu ;\gamma },~~~
\end{eqnarray}%
which are again finite and real energies irrespective of whether the shifts
are taken to be complex or real.

\section{Degenerate complex multi-soliton solutions from Hirota's direct
method}

Hirota's direct method \cite{Hirota} takes a different equivalent form for
nonlinear wave equations as starting point. The system at hand, the KdV
equation (\ref{KdV}), can be converted into Hirota's bilinear form 
\begin{equation}
\left( D_{x}^{4}+D_{x}D_{t}\right) \tau \cdot \tau =0,  \label{Hirota}
\end{equation}%
by means of the variable transformation $u=2(\ln \tau )_{xx}$. The required
combination of Hirota derivatives in terms of ordinary derivatives are%
\begin{eqnarray}
D_{x}^{4}\tau \cdot \tau &=&2\tau _{xxxx}\tau -4\tau _{xxx}\tau _{x}+6\tau
_{xx}\tau _{xx}, \\
D_{x}D_{t}\tau \cdot \tau &=&2\tau _{xt}\tau -2\tau _{x}\tau _{t}.
\end{eqnarray}%
The $\tau $-function can be identified with the Wronskian in the previous
section, up to the ambiguity of an overall factor $\exp \left[
c_{1}x+c_{2}+f(t)\right] $ with arbitrary constants $c_{1}$, $c_{2}$ and
function $f(t)$. Remarkably equation (\ref{Hirota}) can be solved with a
perturbative Ansatz $\tau =\sum\nolimits_{k=0}^{\infty }\varepsilon ^{k}\tau
^{k}$ in an exact manner, meaning that this series terminates at $N$-th
order in $\varepsilon $ for the corresponding $N$-soliton solution. Order by
order one needs to solve the following set of \emph{linear }equations%
\begin{eqnarray}
\left( D_{x}^{4}+D_{x}D_{t}\right) (1\cdot \tau ^{1}+\tau ^{1}\cdot 1)
&=&2(\tau ^{1})_{xt}+2(\tau ^{1})_{xxxx}=0,  \label{zwei} \\
\left( D_{x}^{4}+D_{x}D_{t}\right) (1\cdot \tau ^{2}+\tau ^{1}\cdot \tau
^{1}+\tau ^{2}\cdot 1) &=&0, \\
\left( D_{x}^{4}+D_{x}D_{t}\right) (1\cdot \tau ^{3}+\tau ^{1}\cdot \tau
^{2}+\tau ^{2}\cdot \tau ^{1}+\tau ^{3}\cdot 1) &=&0.
\end{eqnarray}

Let us first see how the Hirota equations are solved using Wronskians
involving Jordan states. We start with the two-soliton solution and take $%
\tau ^{1}=W_{x}\left[ \psi ,\chi ^{(1)}\right] $. Using identity (\ref{wx})
in the form $W_{x}\left[ \psi ,\chi ^{(1)}\right] =c\psi ^{2}$, the first
order Hirota equation (\ref{zwei}) reads%
\begin{equation}
(\tau ^{1})_{xt}+(\tau ^{1})_{xxxx}=(W_{x})_{t}+(W_{x})_{xxx}=c(\psi
^{2})_{t}+c(\psi ^{2})_{xxx}=2c\psi \left( \psi _{t}+\alpha ^{2}\psi
_{x}\right) =0.
\end{equation}%
This equation is solved using the above mentioned nonlinear dispersion
relation, i.e. by taking $\psi (x,t)=\psi (x-\alpha ^{2}t)$.

Next we show how one may carry out the limit to our degenerate solutions
directly on the Hirota multi-soliton solutions. The two-soliton $\tau $%
-function is known to be of the form 
\begin{equation}
\tau _{\alpha ,\beta }(x,t)=1+c_{1}e^{\eta _{a}}+c_{2}e^{\eta _{\beta
}}+c_{1}c_{2}\varkappa (\alpha ,\beta )e^{\eta _{a}+\eta _{\beta }}
\label{series}
\end{equation}%
with $\eta _{a}:=\alpha x-\alpha ^{3}t$ and $\varkappa (\alpha ,\beta ):=$ $%
(\alpha -\beta )^{2}/(\alpha +\beta )^{2}$. Usually the constants $c_{1}$
and $c_{2}$ are set to one. Evidently carrying out the limit \ $\alpha
\rightarrow \beta $ in this variant will simply produce a one-soliton
solution. However, when making use of the freedom to multiply the $\tau $%
-function with an overall factor we define%
\begin{equation}
\tau _{\mu ,\nu ;\alpha ,\beta }(x,t)=\frac{8}{(\alpha -\beta )}e^{\frac{%
\eta _{a}+\eta _{\beta }}{2}+\mu }W\left[ \psi _{\mu +\frac{\alpha -\beta }{%
\alpha +\beta }\nu ,\alpha },\psi _{\mu -\frac{\alpha -\beta }{\alpha +\beta 
}\nu ,\beta }\right] ,  \label{Wt}
\end{equation}%
which produces the series expansion form (\ref{series}) of the $\tau $%
-function with coefficients 
\begin{equation}
c_{1}=-\frac{\alpha +\beta }{\alpha -\beta }e^{\mu +\frac{\alpha -\beta }{%
\alpha +\beta }\nu },\qquad \text{and\qquad }c_{2}=\frac{\alpha +\beta }{%
\alpha -\beta }e^{\mu -\frac{\alpha -\beta }{\alpha +\beta }\nu }.
\label{c12}
\end{equation}%
In this form the limit is easily performed%
\begin{equation}
\tau _{\mu ,\nu ;\alpha ,\alpha }(x,t)=\lim_{\beta \rightarrow \alpha }\tau
_{\mu ,\nu ;\alpha ,\beta }(x,t)=1-2(\alpha x-3\alpha ^{3}t+\nu )e^{\eta
_{a}+\mu }-e^{2\eta _{a}+2\mu }.
\end{equation}%
Since the factor in (\ref{Wt}) has the form of the general ambiguity, the
expression $2\left( \ln \tau _{\mu ,\nu ;\alpha ,\alpha }\right) _{xx}$
produces the same two-soliton solution (\ref{utwo}) as previously obtained.

Similarly, using the identity%
\begin{equation}
\tau _{\mu ,\nu ,\rho ;\alpha ,\beta ,\gamma }(x,t)=\frac{64\exp \left( 
\frac{\eta _{a}+\eta _{\beta }+\eta _{\gamma }+3\mu }{2}\right) }{(\alpha
-\beta )(\alpha -\gamma )(\beta -\gamma )}W\left[ \psi _{\mu +f(\alpha
,\beta ,\gamma ),\alpha },\psi _{\mu +f(\beta ,\gamma ,\alpha ),\beta },\psi
_{\mu +f(\gamma ,\alpha ,\beta ),\gamma }\right] 
\end{equation}%
we obtain the series expansion form (\ref{series}) of the $\tau $-function
for the 3-soliton solution 
\begin{eqnarray}
\tau _{\mu ,\nu ,\rho ;\alpha ,\beta ,\gamma }(x,t) &=&1+c_{1}e^{\eta
_{a}}+c_{2}e^{\eta _{\beta }}+c_{3}e^{\eta _{\gamma }}+c_{1}c_{2}\varkappa
(\alpha ,\beta )e^{\eta _{a}+\eta _{\beta }}+c_{1}c_{3}\varkappa (\alpha
,\gamma )e^{\eta _{a}+\eta _{\gamma }}  \notag \\
&&+c_{2}c_{3}\varkappa (\beta ,\gamma )e^{\eta _{\beta }+\eta _{\gamma
}}+c_{1}c_{2}c_{3}\varkappa (\alpha ,\beta )\varkappa (\alpha ,\gamma
)\varkappa (\beta ,\gamma )e^{\eta _{a}+\eta _{\beta }+\eta _{\gamma }}
\end{eqnarray}%
with coefficients%
\begin{equation}
c_{1}=c(\alpha ,\beta ,\gamma ),\quad c_{2}=c(\beta ,\gamma ,\alpha ),\quad
c_{3}=c(\gamma ,\alpha ,\beta ),~~
\end{equation}%
where%
\begin{equation}
c(x,y,z)=\frac{(x+y)(x+z)}{(x-y)(x-z)}e^{\mu +f(x,y,z)}.
\end{equation}%
Clearly without the information from the previous section it is not obvious
at this stage how to determine the coefficients $c_{i}$ in general,
especially the regularizing shifts.

\section{Degenerate complex multi-soliton solutions from superposition}

It is well-known that the combination of four B\"{a}cklund transformations
combined in a Bianchi-Lamb \cite{Lamb,Bianchi} commutative fashion gives
rise to a \textquotedblleft \emph{nonlinear superposition principle}%
\textquotedblright , e.g. \cite{CenFring}. Introducing the quantity $u=w_{x}$%
, it takes on the form 
\begin{equation}
w_{12}=w_{0}+2\frac{\kappa _{1}-\kappa _{2}}{w_{1}-w_{2}},  \label{super}
\end{equation}%
for the KdV equation where $w_{0}$, $w_{1}$, $w_{2}$ and $w_{12}$ correspond
to different solutions. Relating $w_{1}$ and $w_{2}$ to the standard
one-soliton solution and setting $w_{0}$ to the trivial solution $w_{0}=0$,
the general formula (\ref{super}) becomes%
\begin{equation}
w_{\mu ,\hat{\mu};\alpha ,\beta }=\frac{\alpha ^{2}-\beta ^{2}}{w_{\mu
;\alpha }-w_{\hat{\mu};\beta }},
\end{equation}%
with $w_{\mu ;\alpha }(x,t)=\alpha \tanh \left[ \frac{1}{2}(\alpha x-\alpha
^{3}t+\mu )\right] $, $\kappa _{1}=\alpha ^{2}/2$ and $\kappa _{2}=\beta
^{2}/2$, see \cite{CenFring}. Remarkably in this form the limit $\lim_{\beta
\rightarrow \alpha }w_{\mu ,\hat{\mu};\alpha ,\beta }$ can be performed
directly%
\begin{equation}
\lim_{\beta \rightarrow \alpha }w_{\mu ,\hat{\mu};\alpha ,\beta }=\left\{ 
\begin{array}{cc}
0 & \text{for }\mu \neq \hat{\mu} \\ 
2\alpha \frac{1+\cosh \left( \eta _{\mu ;\alpha }^{(1)}\right) }{\eta
_{0;\alpha }^{(3)}+\sinh \left( \eta _{\mu ;\alpha }^{(1)}\right) }~~~~ & 
\text{for }\mu =\hat{\mu}%
\end{array}%
\right. .
\end{equation}%
The corresponding solution the KdV equation will still be singular, but when
implementing the same shifts as in (\ref{shift2}) we compute 
\begin{equation}
\left( \lim_{\beta \rightarrow \alpha }w_{\mu +\frac{\alpha -\beta }{\alpha
+\beta }\nu ,\mu -\frac{\alpha -\beta }{\alpha +\beta }\nu ;\alpha ,\beta
}\right) _{x}=\lim_{\beta \rightarrow \alpha }\left( w_{\mu +\frac{\alpha
-\beta }{\alpha +\beta }\nu ,\mu -\frac{\alpha -\beta }{\alpha +\beta }\nu
;\alpha ,\beta }\right) _{x}=u_{\mu ,\nu ;\alpha ,\alpha },
\end{equation}%
and thus recover precisely the solution (\ref{utwo}). The relation to the
treatment in section 2 involving DC--transformations is achieved by
considering (\ref{Dar}) for $n=0$ with $V^{(0)}=0$. Then we read off the
identification $w_{\mu ;\alpha }=2\left( \ln \psi _{\mu ,\alpha }\right)
_{x} $, which is confirmed by the explicit expression (\ref{fund}).

Similarly we may carry out the limit on higher soliton solutions. For
instance, iterating (\ref{super}) once more we obtain the three-soliton
solution%
\begin{equation}
w_{\mu ,\nu ,\rho ;\alpha ,\beta ,\gamma }=w_{\mu ;\alpha }+\frac{\beta
^{2}-\gamma ^{2}}{w_{\mu ,\nu ;\alpha ,\beta }-w_{\mu ,\rho ;\alpha ,\gamma }%
},
\end{equation}%
which yields the non-trivial limit%
\begin{equation}
\lim_{\beta ,\gamma \rightarrow \alpha }w_{\mu ,\mu ,\mu ;\alpha ,\beta
,\gamma }=\frac{2\left[ \left[ 1+\left( \eta _{0;\alpha }^{(3)}\right)
^{2}+\cosh \left( \eta _{\mu ;\alpha }^{(1)}\right) \right] \sinh \left( 
\frac{\eta _{\mu ;\alpha }^{(1)}}{2}\right) -\eta _{0;\alpha }^{(9)}\cosh
\left( \frac{\eta _{\mu ;\alpha }^{(1)}}{2}\right) \right] _{x}}{\left[
1+\left( \eta _{0;\alpha }^{(3)}\right) ^{2}+\cosh \left( \eta _{\mu ;\alpha
}^{(1)}\right) \right] \sinh \left( \frac{\eta _{\mu ;\alpha }^{(1)}}{2}%
\right) -\eta _{0;\alpha }^{(9)}\cosh \left( \frac{\eta _{\mu ;\alpha }^{(1)}%
}{2}\right) }.
\end{equation}%
When implementing the appropriate shifts and differentiating once more this
produces precisely the same three-soliton solution as previously constructed
in section 2.2.2.

\section{Conclusions}

We have constructed a novel type of compound soliton solution composed of a
fixed number degenerate one-soliton constituents with the same energy.
Asymptotically, that is for large and small time, the individual
one-solitons travel at the same velocity with almost constant amplitudes. In
the intermediate regime they scatter and exchange their relative position.
Thus the entire collection of one solitons may be viewed as a single
compound object with an internal structure only visible in a certain regime
of time. As we have shown, one may construct solutions in which these
compounds scatter with other (degenerate) multi-solitons at different
velocities.

Technically these compound structures arose from carefully designed limiting
processes of multi-soliton solutions. We have demonstrated how these limits
can be performed within the context of standard techniques of integrable
systems, employing Darboux-Crum transformations involving Jordan states,
Hirota's direct method with specially selected coefficients and on the
nonlinear superposition obtained from B\"{a}cklund transformations. While
the limits led to mathematically admissible nonlinear wave solutions, they
always possess singularities such that their energy becomes infinite. In
order to convert them into physical objects it was crucial to implement in
addition some complex regularizing shifts.

When comparing the different methods, the DC-transformations require the
most substantial modification by the introduction of Jordan states. This
approach is very systematic and the modified transformations always
constitute degenerate soliton solutions. To carry out the limit within the
context of Hirota's direct method requires some guesswork in regards to the
appropriate choice of coefficients, which we overcame here by relying on the
information from the DC-transformations. The nonlinear superposition of
three solutions appears to be the most conductive form for taking the limit
directly. The disadvantage in this approach is that expressions for higher
multi-soliton solutions are rather cumbersome when expressed iteratively. So
far in all approaches the regularizing shift were introduced in a somewhat
ad hoc fashion.

There are various open issues left to be resolved and not reported here.
Evidently the suggested procedure is entirely generic and not limited to the
KdV equations or the particular type of solutions and boundary conditions
considered here \cite{arancibia2015chiral}. It would be interesting to apply
them to other types of integrable systems as that might help to unravel some
further universal features. For instance, one expects that the regularizing
shifts can be cast into a more universal form that might be valid for any
arbitrary number of degeneracies when exploiting further their ambiguities.
Furthermore it is desirable to complete the argument on why the energies of
these complex solutions are real. This follows immediately when they and the
corresponding Hamiltonians are $\mathcal{PT}$-symmetric. As demonstrated in 
\cite{CenFring}, this can be achieved with suitable real shifts in time or
space, but in addition one also requires the model to be integrable. We
report on these issues in more detail elsewhere \cite{CenCorreaFring}.

\medskip

\noindent \textbf{Acknowledgments:} FC would like to thank the Alexander von
Humboldt Foundation (grant number CHL 1153844 STP) for financial support and
City University London for kind hospitality.

%%\bibliographystyle{phreport}
%%\bibliography{acompat,Ref}

\end{document}